%% file: DPF2013_Oblath.tex
\documentclass[12pt]{article}
\usepackage{graphicx}

\newcommand\pubnumber{DPF2013-187}
\newcommand\pubdate{\today}

\def\support{\footnote{Presented on behalf of the Project 8 Collaboration.}}

\def\Title#1{\begin{center} {\Large #1 } \end{center}}
\def\Author#1{\begin{center}{ \sc #1} \end{center}}
\def\Address#1{\begin{center}{ \it #1} \end{center}}

\newcommand\pubblock{\rightline{\begin{tabular}{l} \pubnumber\\
         \pubdate  \end{tabular}}}
\newenvironment{Abstract}{\begin{quotation}  }{\end{quotation}}
\newenvironment{Presented}{\begin{quotation} \begin{center} 
             PRESENTED AT\end{center}\bigskip 
      \begin{center}\begin{large}}{\end{large}\end{center} \end{quotation}}

\input econfmacros.tex

\begin{document}
\begin{titlepage}
\pubblock

\vfill
\Title{Project 8: Using Radio-Frequency Techniques to Measure Neutrino Mass}
\vfill
\Author{ N.\,S.~Oblath\support}
\Address{Laboratory for Nuclear Science\\Massachusetts Institute of Technology\\Cambridge, MA}
\vfill
\begin{Abstract}
The Project 8 experiment aims to measure the neutrino mass using tritium beta decays.  Beta-decay electron energies will be measured with a novel technique: as the electrons travel in a uniform magnetic field their cyclotron radiation will be detected. The frequency of each electron's cyclotron radiation is inversely proportional to its total relativistic energy; therefore, by observing the cyclotron radiation we can make a precise measurement of the electron energies.  The advantages of this technique include scalability, excellent energy resolution, and low backgrounds.  The collaboration is using a prototype experiment to study the feasibility of the technique with a $^{83m}$Kr source.  Demonstrating the ability to see the 17.8~keV and 30.2~keV conversion electrons from $^{83m}$Kr will show that it may be possible to measure tritium beta-decay electron energies ($Q \approx 18.6$~keV) with their cyclotron radiation.  Progress on the prototype, analysis and signal-extraction techniques, and an estimate of the potential future of the experiment will be discussed.
\end{Abstract}
\vfill
\begin{Presented}
DPF 2013\\
The Meeting of the American Physical Society\\
Division of Particles and Fields\\
Santa Cruz, California, August 13--17, 2013\\
\end{Presented}
\vfill
\end{titlepage}
\def\thefootnote{\fnsymbol{footnote}}
\setcounter{footnote}{0}

\section{Neutrino Mass via Tritium Beta Decay}

While neutrino oscillation experiments have successfully shown that neutrinos change flavor, and therefore have non-zero mass, the absolute mass scale remains unknown.  The simplest way to directly measure the mass of the neutrino is using beta decays.  Neutrino mass has an effect on the kinematics of decay process~\cite{ref:tritiumbetadecay}.  While the neutrinos themselves are difficult to measure, the energies of the outgoing electrons can be precisely determined.  The neutrino mass can then be inferred from the shape of the electron energy spectrum:

\begin{equation}
\frac{dN}{dK_e} \propto F(Z,K_{e}) \cdot p_e \cdot (K_{e}+m_{e}) \cdot (E_{0}-K_{e}) \cdot \sum_{i=1}^{3}|U_{ei}|^2 \sqrt{(E_0-K_e)^2-m_i^2} \cdot \Theta(E_0-K_e-m_i).
\label{eq:betadecayspectrum}
\end{equation}

The Fermi function, $F(Z,K_{e})$, takes into account the Coulomb interactions of the electron with the recoiling nucleus; $Z$ is the proton number of the final-state nucleus, $K_e$ is the electron's kinetic energy, $p_e$ is the electron's momentum, $E_0$ is the Q-value of the decay, and $U_{ei}$ are the elements of the PMNS matrix for neutrino mass states $m_{i,\,i=1-3}$.  The only dependence on the neutrino mass comes from the phase-space factor.  The shape of the spectrum is independent of all other properties of the neutrino, including whether neutrinos are Majorana or Dirac particles.

One technique being used to precisely measure the beta-decay spectrum relies on a spectrometer to precisely select high-energy electrons from tritium decays.  The most recent experiments to use this technique are the Mainz and Troitsk experiments.  They placed similar limits on the neutrino mass: $m_{\beta\nu} < 2.3\ \mathrm{eV}$~\cite{ref:mainz,ref:troitsk}.  KATRIN, the next-generation of spectrometer-type experiments, aims to lower that limit by an order of magnitude, to 200~meV (90\% CL)~\cite{ref:katrin}.  KATRIN is currently under construction and commissioning in Karlsruhe, Germany.

The lower limits for the neutrino mass from oscillation experiments provide a strong motivation for probing to lower neutrino masses.  However, with KATRIN, the technologies used in spectrometer-type tritium experiments have been pushed to their current practical limits.  A new technique is needed to push the mass sensitivity lower. 

\section{A New Technique}

The Project 8 collaboration proposes an alternate method of measuring the electron energies: measure the cyclotron radiation emitted by the electrons spiraling around magnetic field lines.  An enclosed volume of tritium is placed in a uniform magnetic field, and as the tritium nuclei decay, the electrons will spiral around the magnetic fields lines.  The spiraling electrons are being accelerated, and therefore emit cyclotron radiation.  The frequency of that radiation is proportional to the magnetic field strength, and inversely proportional to the electron's kinetic energy:

\begin{equation}
\omega = \frac{eB}{\gamma m_e} = \frac{\omega_c}{\gamma} = \frac{\omega_c}{1+K_e/(m_e c^2)}.
\label{eq:frequency}
\end{equation}

By measuring the frequency of the cyclotron radiation, one can measure the electron's kinetic energy without interfering with the electron itself.  Using a 1-T magnetic field, the endpoint of the tritium spectrum (18.6-keV) falls around 26~GHz.  The power emitted as cyclotron radiation depends both on the relativistic velocity electron, $\beta$, and the angle at which it is emitted relative to the direction of the magnetic field, $\theta$:

\begin{equation}
P(\beta,\theta) = \frac{1}{4\pi\epsilon_0}\frac{2 q^2 \omega_c^2}{3c}\frac{\beta_{\perp}^2}{1-\beta^2}, \qquad \beta_{\perp} \equiv \beta \sin{\theta}.
\label{eq:power}
\end{equation}

The electrons that radiate the most power will be the easiest to detect, because the signal-to-noise ratio will be higher.  Equation~\ref{eq:power} shows that the power will be greatest for electrons with $\theta \approx 90^{\circ}$.  Conveniently, these electrons also travel the slowest along the magnetic field lines, increasing the amount of time they can be observed.

\begin{figure}
\centering
\includegraphics[width=0.45\textwidth]{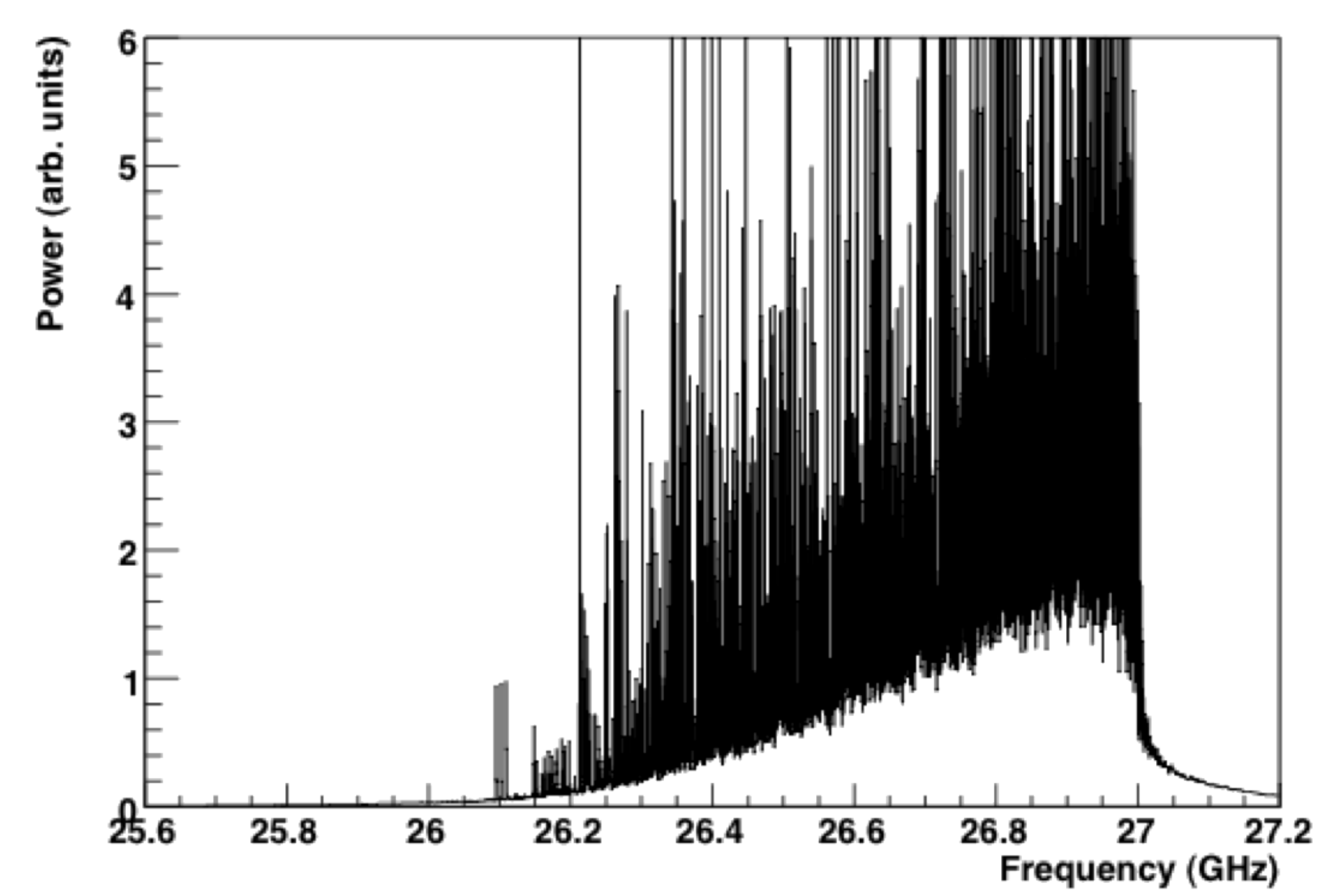}
\includegraphics[width=0.45\textwidth]{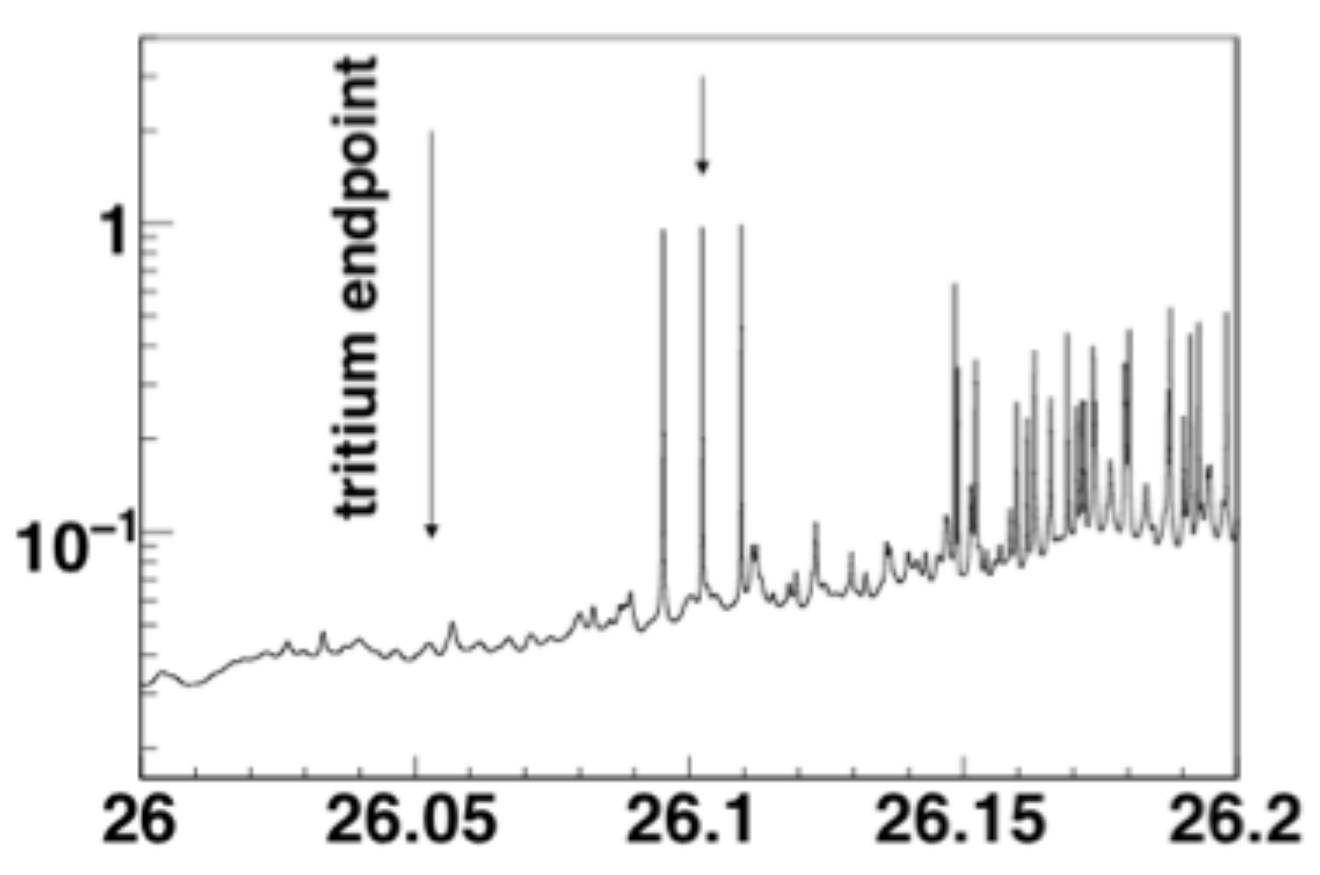}
\caption{\label{fig:powerspectrum}Simulated power spectrum from the hypothetical experiment described in~\cite{ref:formaggio_monreal_2009}.  $10^5$ beta decays were simulated over 30~$\mu$s.  The vertical arrow indicates the location of the tritium beta-decay endpoint in frequency space. The full spectrum is on the left, and the endpoint region is on the right.  The triplet of peaks from a high-energy electron are easily distinguished from the background.}
\end{figure}

For the hypothetical experiment described in~\cite{ref:formaggio_monreal_2009}, the simulated power spectrum measured from $10^5$ tritium decays in 30~$\mu$s is shown in Fig.~\ref{fig:powerspectrum}~(left).  Since frequency is inversely proportional to electron energy, the rare high-energy electrons are at lower frequency, near 26~GHz.  Low-energy electrons, making up the vast majority of the spectrum, are piled up towards 27~GHz.  The pileup from low-energy electrons is a possible complication: unlike spectrometer-type experiments, we do not have an intrinsic method for rejecting low-energy electrons.  Instead, we can narrow the bandwidth such that the event rate is low enough that individual events can be identified.

The primary concern for making a precise electron energy measurement is the ability to measure frequency precisely.  The desired energy precision is therefore the place to start in considering the requirements for this type of experiment.  To achieve the necessary energy precision, $\Delta E$, we need to achieve a relative frequency precision of $\Delta f / f = \Delta E / m_e$.  KATRIN is designed to achieve $\Delta E \approx 1\ \mathrm{eV}$; for Project 8 to achieve a similar accuracy means that $\Delta f / f \approx 2 \times 10^{-6}$.  This accuracy is reasonable with current technologies.  With a 1-T magnetic field, $\Delta f \approx 52\ \mathrm{kHz}$ at 26~GHz.

The desired frequency accuracy determines for how long we must be able to observe single electrons.  To have a frequency resolution of $\Delta f$, we must measure each electron for $t_{\mathrm{min}} = 1 / \Delta f$.  With the design parameters discussed above, the electrons must be coherently measured for at least 20~$\mu$s.  The minimum measurement time places constraints on a number of physical parameters of the experiment.  The gas density must be low enough that, on average, 18.6~keV electrons can travel for $t_{\mathrm{min}}$ without scattering.  Furthermore, the experiment must be large enough so that the electron can be tracked continuously.

The signal detected for a single electron may be more complicated than the single frequency at which the cyclotron radiation is emitted.  In particular, the detected signal can include a Doppler shift due to the velocity of the electron parallel to the magnetic field, $\beta_{\parallel}$, a dependence on the electron-antenna distance, and effects from the angular dependence of the power distribution of the radiation.  The way these effects are represented in the data depends strongly on the antenna configuration.  For the hypothetical experiment described in~\cite{ref:formaggio_monreal_2009}, if the signals from the different antennas are summed coherently, there will be sideband peaks from the Doppler shift.  Fig.~\ref{fig:powerspectrum}~(right) zooms in on the high-energy (low-frequency) region of the power spectrum shown previously.  In this simulation a triplet of peaks from a single high-energy electron are easily distinguishable.  Though it is an antenna-design-dependent effect, the triplet of peaks due to the Doppler shift could be a convenient tool for tagging electrons.

\subsection{Prototype Experiment}

The Project 8 Collaboration has put together a prototype experiment to explore the practical use of cyclotron radiation as a method for measuring the decay-electron energy.  The initial goal of the prototype is to verify that we can, in fact, detect the cyclotron radiation from a single electron.  We will use a $^{83m}$Kr radioactive source, which emits a monoenergetic electron.  This excited nucleus emits 17.8~keV or 30~keV electrons, and has a half-life of 1.83 hours.  The source is a good stand-in for tritium: it is gaseous, emitting the electrons isotropically, and the energy of one of the decay branches is close to the tritium-decay endpoint.

\begin{figure}
\centering
\includegraphics[width=5in]{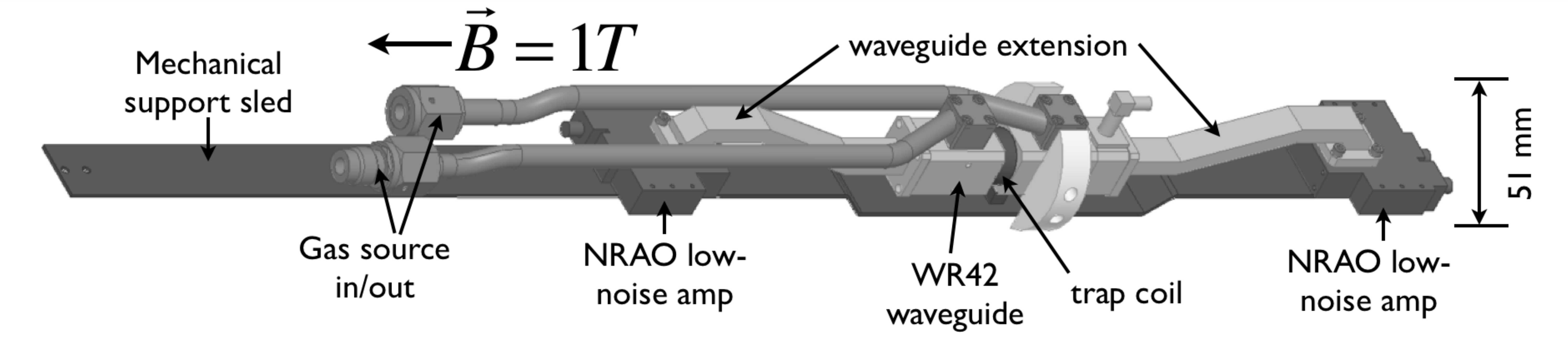}
\caption{\label{fig:prototype}Diagram of the magnet-bore insert for the prototype experiment located at the University of Washington, in Seattle, WA.  The configuration shown was used to take data in January, 2013.}
\end{figure}

Figure~\ref{fig:prototype} shows a diagram of the magnet insert for the prototype experiment, which is located at the University of Washington, in Seattle, WA.  A superconducting solenoid provides the 1-T magnetic field.  The electrons are trapped in a small ($\approx 1 \mathrm{mm}^3$) magnetic bottle in the bore of the magnet.  The magnetic field from the solenoid traps the electrons in the horizontal plane; a trapping coil within the bore of the magnet decreases the field slightly in a small volume, trapping the electrons vertically.  Whether or not electrons are trapped depends on the depth of the magnetic bottle potential, and the pitch angle of the electrons.  Electrons with large $\beta_{\perp}$ will be trapped.  Fortunately, these electrons also emit the most power as cyclotron radiation.  Only electrons with large pitch angles ($\theta \ge 85^{\circ}$) are trapped.  Though this angle selection severely limits the number of electrons we will detect, it maximizes the signal-to-noise ratio.

The ability to open the magnetic bottle by turning off or reversing the current in the trapping coil will allow us to confirm that we are indeed trapping electrons, and accurately measure the noise levels. In addition to detecting the cyclotron radiation, we will employ more traditional means of detecting electrons to monitor the presence of $^{83m}$Kr in the trap, and verify that electrons are actually trapped.

\begin{figure}
\centering
\includegraphics[width=0.45\textwidth]{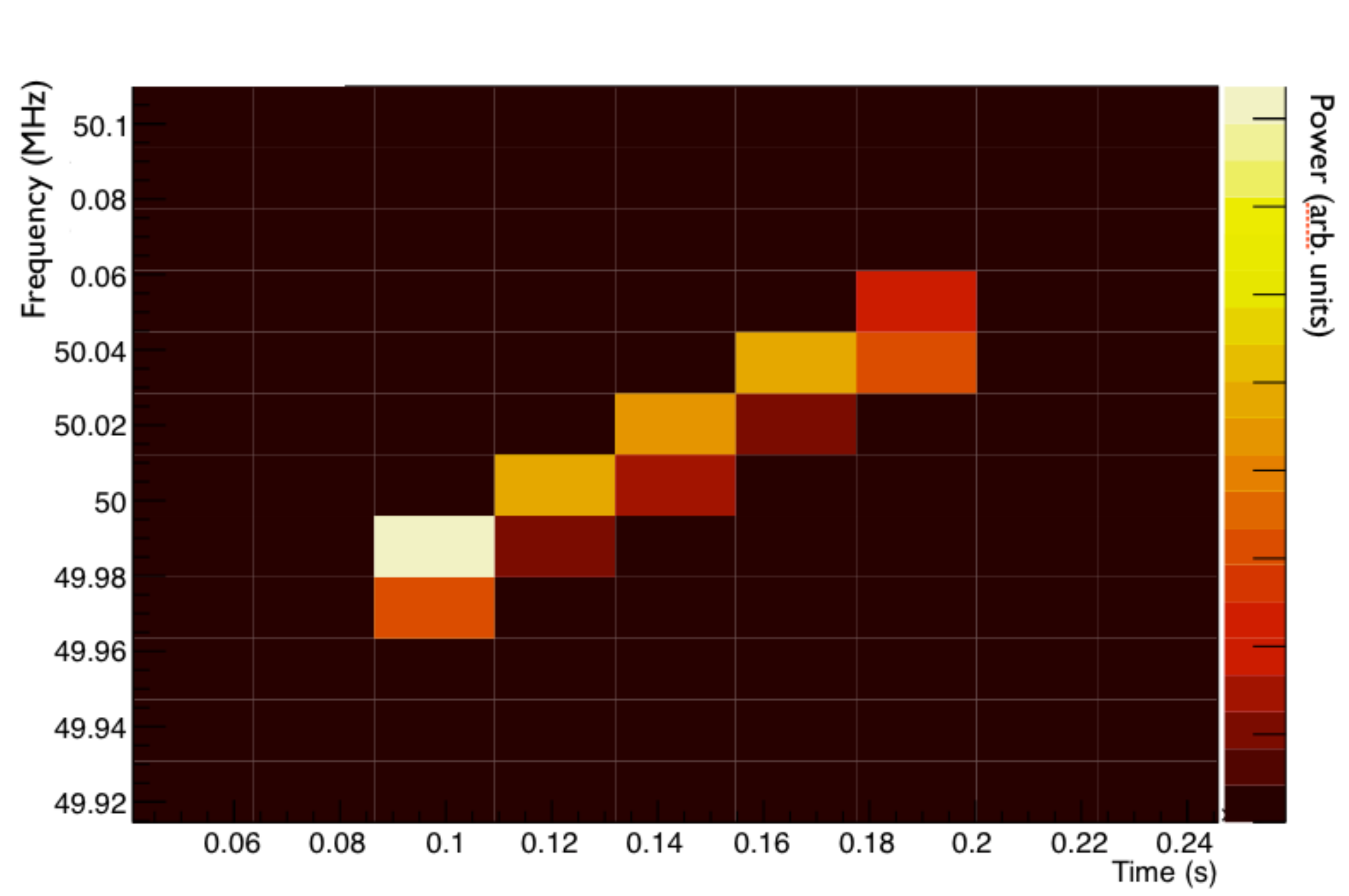}
\includegraphics[width=0.45\textwidth]{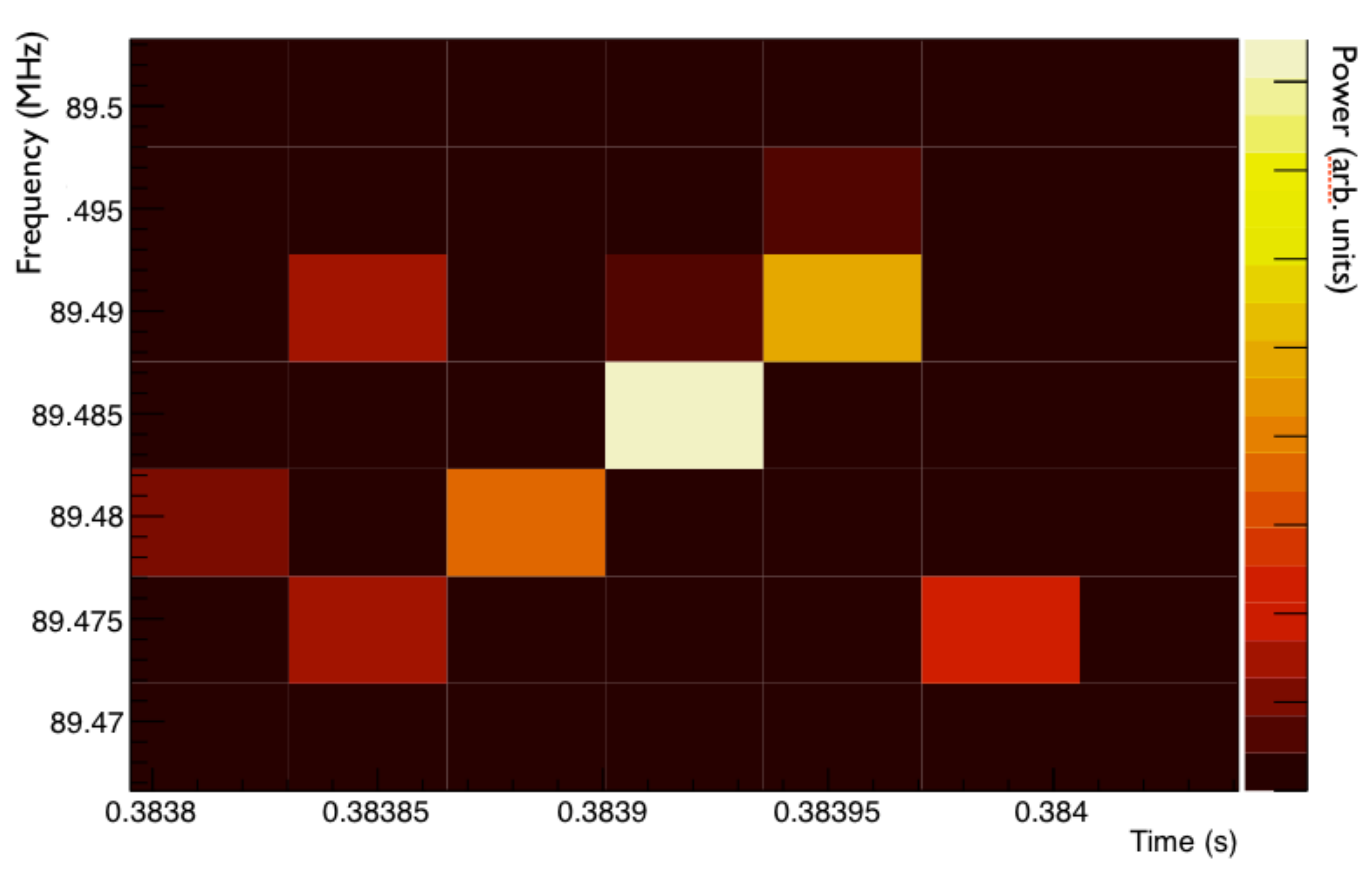}
\caption{\label{fig:chirp}The time-frequency-space representation of our data.  On the left is a simulated chirp.  The candidate on the right resembles what we expect our signal to look like, but is actually a random noise fluctuation from a run in which the trapping-magnet current was reversed, so no electrons were being stored in the magnetic bottle.}
\end{figure}

The cyclotron radiation is detected with a waveguide coupled to two low-noise cryogenic amplifiers.  The rectangular cavity of the waveguide also serves to contain the $^{83m}$Kr gas.  The signals from the amplifier are mixed down to lower frequencies, digitized and written to disk.  After the data has been recorded, we analyze it to search for excesses of power as a function of frequency.  Fig.~\ref{fig:chirp} (left) shows a simulated ``chirp'' signal in time-frequency space.  Each column is created by taking a Fourier Transform of a $\approx 40$-ms time slice.  The signal rises in frequency as a function of time because the electron loses energy to the cyclotron radiation.  The main background in our analysis is the random noise that sometimes fluctuates high enough to mimic a signal.  As Fig.~\ref{fig:chirp} (right) shows, the noise can form clusters that look like signal chirps.  We can identify electron candidates by finding a peak of candidate chirps as a function of frequency; using the clustering of high-power bins as a function of time allows us to significantly reduce the background candidate rate.

Data was taken in January, 2013, though we have not yet seen any indication of trapped electrons in that data set.  We are currently developing more sensitive analysis techniques, and plan on taking data in Fall 2013 after making several improvements to the apparatus.

\section{Future Work}

Once we have shown that we can detect single electrons using their cyclotron radiation, we will investigate the energy resolution achievable with this type of setup.  The $^{83m}$Kr source is particularly useful for this purpose, since the electrons are monoenergetic.

Finally, we want to demonstrate that the signal from cyclotron radiation can be used to identify electrons and determine their energy without additional detection methods.  Though for the initial stages of the prototype the data acquisition is untriggered, we will need to develop the ability to recognize electrons and trigger the recording of data.

 This research is supported in part by DOE grant DE-FG02-97ER41020 and the National Science Foundation.

\end{document}

%% file: econfmacros.tex
\def\beq{\begin{equation}}
\def\eeq#1{\label{#1}\end{equation}}
\def\eeqn{\end{equation}}

\def\beqa{\begin{eqnarray}}
\def\eeqa#1{\label{#1}\end{eqnarray}}
\def\eeqan{\end{eqnarray}}

\let\bar=\overbar

\def\Dslash{\not{\hbox{\kern-4pt $D$}}}
\def\dslash{\not{\hbox{\kern-2pt $\del$}}}

\def\msb{{\bar{\ssstyle M \kern -1pt S}}}

